\documentclass{aa}  
\usepackage{graphicx}
\usepackage{graphics}
\usepackage{txfonts}
\usepackage{placeins}
\usepackage[]{hyperref}
\hypersetup{colorlinks=false}
\usepackage[utf8]{inputenc}
\usepackage[english]{babel}
\usepackage{url}
\usepackage{amssymb}
\usepackage{color}
\usepackage{multirow}
\usepackage{todonotes}
\usepackage{soul}
\usepackage{ulem}
\usepackage{natbib}

\newcommand{\rev}[1]{{{#1}}}

\newcommand{\emm}[1]{\ensuremath{#1}}
\newcommand{\emr}[1]{\emm{\mathrm{#1}}}

\newcommand{\unit}[1]{\emr{\,#1}}
\newcommand{\kms}{\unit{km\,s^{-1}}}

  \newcommand{\Tabobs}{
  \begin{table*}
    \centering %
    \caption{Observation parameters.}
    \begin{tabular}{lcccccc}
  \hline \hline
 Name & RA & DEC & Telescope & Observing mode & Spectral resolution & Int. Time \\
            &  J2000    &    J2000      &                     &                       & (kHz)                      & (hours) \\
  \hline                                                          
  BL Lac &    22:02:43.29 & 42:16:39.9        & Yebes 40m & Position Switching$^a$         &          38                         &  5.3\\
  BL Lac &      22:02:43.29 & 42:16:39.9          & Yebes 40m & Frequency Switching         &             38                      & 78.9  \\
    \hline
  \end{tabular} 
   \tablefoot{
  \tablefoottext{a}{The reference position is located at equatorial coordinates (J2000) RA=$22^h02^m36^s$ and DEC=$+42^{\circ}15'30''$.}
  }
    \label{tab:obs}
  \end{table*}
  }

  \newcommand{\TabLine}{
  \begin{table*}[h!]
    \centering 
    \caption{Detected lines in the Q band.}
        \begin{tabular*}{18cm}{lccccccc}
  \hline \hline
 Molecule & Transition$^a$ & Frequency$^a$ & rms$^b$ &    $\tau$ & $V$ & FWHM & $\int \tau dv$  \\
                &                       & (MHz)             &   (10$^{-3}$)        & (10$^{-2}$)       & (\kms) & (\kms) & (10$^{-2}$\kms)   \\
                                &                       &                  &         &        & &  &    \\

  \hline
  CS    &       $1-0$   &       48990.9549      &       2.10    &       6.20    &       $-0.65  \pm     0.04$   &       $0.86   \pm     0.06    $&      $5.60   \pm     0.500$ \\
CS      &       $1-0$   &       48990.9549      &       0.00    &       17.40&  $       -1.52   \pm     0.04$   &       $0.76   \pm     0.02$   &       $14.20  \pm     0.500$ \\                      C$^{34}$S       &       $1-0$   &       48206.9411      &       0.87    &       0.19    &       $-0.40  \pm     0.14$   &       $0.69   \pm     0.40$   &       $0.14   \pm     0.070$\\
C$^{34}$S       &       $1-0$   &       48206.9411      &       0.87    &       0.69    &       $-1.42  \pm     0.05$   &       $0.80   \pm     0.12$   &         $0.58   \pm     0.070$\\
                                                                                                                                                                                                
$^{13}$CS       &       $1-0$   &       46247.5632      &       0.70    &       0.19    &       $-1.17  \pm     0.11$   &       $0.88   \pm     0.24$   &       $0.18   \pm     0.042$\\
                                                                                                                                                                                                
HCS$^+$ &       $1-0$   &       42674.1954      &       0.40    &       0.38    &       $-0.57  \pm     0.16$   &       $1.04   \pm     0.22$   &       $0.42   \pm     0.130$\\
HCS$^+$ &       $1-0$   &       42674.1954      &       0.40    &       0.20    &       $-1.39  \pm     0.21$   &       $0.76   \pm     0.26$   &       $0.16   \pm     0.130$\\ 
                                                                                                                                                                                                
CCS     &       $2_3-1_2$       &       33751.3696      &       0.24    &       0.14    &       $-1.44  \pm     0.09$   &       $1.09   \pm     0.12$   &       $0.16   \pm     0.020$\\ 
CCS     &       $2_3-1_2$       &       33751.3696      &       0.24    &       0.06    &       $-0.38  \pm     0.19$   &       $1.57   \pm     0.13$   &       $0.09   \pm     0.020$\\
CCS     &       $3_4-2_3$       &       45379.0460      &       0.53    &       0.18    &       $-1.20  \pm     0.10$   &       $0.82   \pm     0.21$   &       $0.16   \pm     0.035$\\        
                                                                                                                                                                                                
p-H$_2$CS       &       $1_{01}-0_{00}$ &       34351.4300      &       0.21    &       0.04    &       $-0.58  \pm     0.20$   &       $2.06   \pm     0.32$   &       $0.10   \pm     0.020$ \\
                                                                                                                                                                                                
l-C$_3$H        &       $3/2_1 - 1/2_1  f$      &       32617.0156      &       0.30    &       0.12    &       $-0.90  \pm     0.17$   &       $1.32   \pm     0.30$   &       $0.17   \pm     0.040$ \\  
l-C$_3$H        &       $3/2_2 - 1/2_1  f$      &       32627.2970      &       0.30    &       0.41    &       $-0.73  \pm     0.03$   &       $1.53   \pm     0.08$   &       $0.67   \pm     0.031$\\ 
l-C$_3$H        &       $3/2_1 - 1/2_0  f$      &       32634.3890      &       0.29    &       0.14    &       $-0.76  \pm     0.10$   &       $1.60   \pm     0.16$   &       $0.25   \pm     0.030$\\ 
l-C$_3$H        &       $3/2_2 - 1/2_1 e$       &       32660.6450      &       0.30    &       0.44    &       $-0.88  \pm     0.03$   &       $1.43   \pm     0.07$   &       $0.67   \pm     0.030$\\ 
l-C$_3$H        &       $3/2_1 - 1/2_0  e$      &       32663.3610      &       0.26    &       0.14    &       $-0.84  \pm     0.14$   &       $2.24   \pm     0.40$   &       $0.33   \pm     0.050$\\ 
l-C$_3$H        &       $3/2_1 - 1/2_1  e$      &       32667.6680      &       0.26    &       0.09    &       $-0.93  \pm     0.13$   &       $1.05   \pm     0.24$   &       $0.09   \pm     0.022$\\        
                                                                                                                                                                                                
C$_3$H$^+$      &       $2-1$   &       44979.5486      &       0.50    &       0.12    &       $-1.50  \pm     0.13$   &       $0.58   \pm     0.28$   &       $0.07   \pm     0.060$ \\ 
C$_3$H$^+$      &       $2-1$   &       44979.5486      &       0.50    &       0.21    &       $-0.55  \pm     0.16$   &       $1.30   \pm     0.36$   &       $0.29   \pm     0.070$ \\ 
                                                                                                                                                                                                
o-l-C$_3$H$_2$  &       $2_{12}-1_{11}$ &       41198.3354      &       0.32    &       0.41    &       $-0.93  \pm     0.04$   &       $1.46   \pm     0.08$   &       $0.63   \pm     0.034$ \\
o-l-C$_3$H$_2$  &       $2_{11}-1_{10}$ &       41967.6705      &       0.40    &       0.57    &       $-0.94  \pm     0.03$   &       $1.45   \pm     0.07$   &       $0.87   \pm     0.037$ \\      
p-l-C$_3$H$_2$  &       $2_{02}-1_{01}$ &       41584.6755      &       0.43    &       0.35    &       $-0.93  \pm     0.06$   &       $1.65   \pm     0.12$   &       $0.61   \pm     0.040$ \\ 
                                                                                                                                                                                                
p-c-C$_3$H$_2$  &       $2_{11}-2_{02}$ &       46755.6100      &       0.70    &       0.35    &       $-1.35  \pm     0.10$   &       $0.74   \pm     0.17$   &       $0.28   \pm     0.110$ \\ 
p-c-C$_3$H$_2$  &       $2_{11}-2_{02}$ &       46755.6100      &       0.70    &       0.22    &       $-0.47  \pm     0.28$   &       $1.39   \pm     0.34$   &       $0.33   \pm     0.120$ \\

C$_4$H  &       $4_{9/2} - 3_{7/2}$     &       38049.6912      &       0.30    &       0.48    &       $-0.66  \pm     0.03$   &       $1.58   \pm     0.07$   &       $0.81   \pm     0.030$ \\ 
C$_4$H  &       $4_{7/2} - 3_{5/2}$     &       38088.4805      &       0.32    &       0.42    &       $-1.05  \pm     0.03$   &       $1.38   \pm     0.07$   &       $0.62   \pm     0.029$ \\      
C$_4$H  &       $5_{11/2} - 4_{9/2}$    &       47566.8139      &       0.76    &       0.41    &       $-0.76  \pm     0.08$   &       $1.50   \pm     0.16$   &       $0.66   \pm     0.061$ \\      
C$_4$H  &       $5_{9/2} - 4_{7/2}$     &       47605.4883      &       0.72    &       0.44    &       $-1.20  \pm     0.06$   &       $1.25   \pm     0.13$   &       $0.59   \pm     0.056$ \\ 
                                                                                                                                                                                                
p-H$_2$CCO      &       $2_{02}-1_{01}$         &       40417.9504      &       0.37    &       0.11    &       $-0.92  \pm     0.13    $&$     1.10    \pm     0.29    $&      $0.13   \pm     0.030$ \\ 
o-H$_2$CCO      &       $2_{12}-1_{11}$  &      40039.0220      &       0.28    &       0.15    &       $-0.99  \pm     0.08    $&$     1.26    \pm     0.16    $&      $0.20   \pm     0.024$ \\      
o-H$_2$CCO      &       $2_{11}-1_{10}$  &      40793.8320      &       0.36    &       0.17    &       $-1.20  \pm     0.08    $&$     1.10    \pm     0.18    $&      $0.20   \pm     0.029$ \\      
                                                                                                                                                                                                
E-CH$_3$CHO     &       $2_{02}-1_{01}$ &       38506.0348      &       0.32    &       0.31    &       $-1.08  \pm     0.05$   &       $1.36   \pm     0.10$   &       $0.45         \pm     0.030$\\ 
A-CH$_3$CHO     &       $2_{02}-1_{01}$ &       38512.0786      &       0.32    &       0.30    &       $-1.10  \pm     0.05$   &       $1.52   \pm     0.11$   &       $0.48   \pm     0.030$\\        
E-CH$_3$CHO     &       $2_{11}-1_{10}$  &      39362.5367      &       0.33    &       0.13    &       $-0.99  \pm     0.11$   &       $1.26   \pm     0.21$   &       $0.18   \pm     0.030$\\        
A-CH$_3$CHO     &       $2_{11}-1_{10}$ &       39594.2893      &       0.32    &       0.14    &       $-0.99  \pm     0.10$   &       $1.36   \pm     0.20$   &       $0.20   \pm     0.030$\\        
E-CH$_3$CHO     &       $2_{12}-1_{11}$  &      37686.9320      &       0.22    &       0.10    &       $-1.10  \pm     0.10$   &       $1.23   \pm     0.21$   &       $0.13   \pm     0.020$\\        
A-CH$_3$CHO     &       $2_{12}-1_{11}$ &       37464.2044      &       0.27    &       0.09    &       $-1.05  \pm     0.12$   &       $1.14   \pm     0.29$   &       $0.10   \pm     0.024$\\        
                                                                                                                                                                                
A-CH$_3$CN      &       $2_0-1_0$       &       36795.5700      &       0.30    &       0.18    &       $-0.99  \pm     0.06$   &       $1.47   \pm     0.13$   &       $0.29   \pm     0.028$  \\
E-CH$_3$CN      &       $2_1-1_1$       &       36794.7652      &       0.30    &       0.21    &       $3.17   \pm     0.06$   &       $1.09   \pm     0.10$   &       $0.25   \pm     0.025$ \\ 
                                                                                                                                                                                                
HC$_3$N &       $4-3$   &       36392.3240      &       0.23    &       0.08    &       $-0.94  \pm     0.13$   &       $2.02   \pm     0.45$   &       $0.18   \pm     0.030$\\ 
HC$_3$N &       $5-4$   &       45490.3138      &       0.50    &       0.20    &       $-1.14  \pm     0.10$   &       $1.38   \pm     0.21$   &       $0.29   \pm     0.040$ \\                                                                                                                                                                                              
                                                                                                                                                                                                
HNCO    &       $2_{02}-1_{01}$ &       43963.0395      &       0.44    &       0.10    &       $-0.76  \pm     0.19$   &       $1.47   \pm     0.43$   &       $0.16   \pm     0.039$ \\ 
                                                                                                                                                                                        
SiO     &       $1-0$   &       43423.8530      &       0.35    &       0.31    &       $-0.25  \pm     0.05$   &       $1.47   \pm     0.12$   &       $0.48   \pm     0.030$ \\ 

  \hline
  \end{tabular*} 
   \tablefoot{
   \tablefoottext{a}{Spectroscopic data are taken from the Cologne Database for Molecular Spectroscopy (CDMS) \citep{Muller:2005,Endres:2016} and from the JPL catalog \citep{Pickett:1998}.}
\tablefoottext {b}{rms on the line to continuum determined on the normalized spectra.}
   }
    \label{tab:lines}
  \end{table*}
  }

\newcommand{\Tabupper}{
  \begin{table*}[h!]
    \centering 
    \caption{Upper limits for the selected lines in the Q band.}
    \begin{tabular*}{17cm}{lccccccc}
  \hline \hline
  Molecule & Transition$^a$ & Frequency$^a$ & rms$^b$ &  $\int \tau dv$$^c$  & $N/\int \tau dv$ &$N$ & X$^f$ \\
               &                       & (MHz)                  &   (10$^{-3}$)           & (10$^{-2}$\kms)   & (10$^{13}$  cm$^{-2}$/\kms) & (10$^{11}$cm$^{-2}$ & (10$^{-10}$)\\
  \hline
CCO & $2_3 - 1_2$ & 45826.706 & 0.55 & $<0.1$ &  16$^d$ & $<1.6$$^d$ &$<1.9$\\
CCN &$ 3/2_{5/2}-1/2_{3/2}$ & 35422.684 & 0.25 &$<0.05$ & 447$^d$& $<23$$^d$ &$<27$ \\
 C$_4$H$^-$ & $4-3$ & 37239.402 & 0.24& $<0.05$& 0.67$^d$ & $<0.03$$^d$  &$<0.04$\\
CH$_3$CCH-A & $2_0-1_0$ & 34183.414 & 0.33 & $<0.07$ & 37.6$^d$ & $<2.64$$^d$ & $<3.1$\\ 
C$_3$N &  $4_{9/2} - 3_{7/2}$ & 39571.397 & 0.31  & $<0.06$ & 13.3$^d$ & $<0.8$$^d$ &$<0.95$\\
 C$_6$H & $25/2-23/2$ &  34654.029 & 0.23& $<0.05$ & 7.85$^d$ & $<0.4$$^d$ &$<0.5$\\ 
  \hline
  \end{tabular*} 
   \tablefoot{
   \tablefoottext{a}{Spectroscopic data are taken from the Cologne Database for Molecular Spectroscopy (CDMS) \citep{Muller:2005,Endres:2016} and from the JPL catalog \citep{Pickett:1998}.} 
\tablefoottext {b}{rms on the line to continuum determined on the normalized spectra.}
\tablefoottext {c}{Assuming a Gaussian absorption profile with FWHM = 1.5 \kms.}
\tablefoottext   {d}{Assuming an excitation temperature of 5~K}
\tablefoottext   {e}{Abundance relative to H$_2$ using N(H$_2$)=$8.6 \times 10^{20}$~cm$^{-2}$.}
   }
    \label{tab:upper}
  \end{table*}
  }

\newcommand{\Tabcol}{
  \begin{table*}
    \centering 
    \caption{Column densities and abundances}
\begin{tabular}{lcccc}
\hline \hline
Species & Line & $N/\int \tau dv$$^a$                                    & $N$$^a$ & X$^b$ \\
            &  &       (10$^{13}$ cm$^{-2}$/\kms) &          (10$^{11}$  cm$^{-2}$) & (10$^{-10}$)\\
\hline
CS  &           $1-0$  & 1.30 &         $26  \pm 3$ &    $30 \pm 3 $ \\
C$^{34}$S & $1-0$ & 1.33 &       $0.96 \pm 0.1$ & $1.1 \pm 0.1$ \\
$^{13}$CS & $1-0$ & 1.42 &     $0.25 \pm 0.06$ & $0.29 \pm 0.07 $ \\
HCS$^+$ &  $1-0$ &  1.62 &      $0.94 \pm 0.2 $ & $1.1\pm 0.25$\\
CCS &   $2_3 - 1_2$ &  1.87 &  $0.47 \pm 0.1$ & $0.5 \pm 0.1$ \\
p-H$_2$CS & $1_{01}-0_{00} $ & 3.26 & $0.31\pm 0.07  $ & $0.37 \pm 0.08$ \\
l-C$_3$H &   $3/2-1/2 $  & 2.92 &          $2.0 \pm 0.1$ & $2.3 \pm 0.2$\\
C$_3$H$^+$ & $2-1$ & 0.82 &       $0.29 \pm 0.06 $  & $0.34 \pm 0.07 $\\
o-l-C$_3$H$_2$  &       $2_{12}-1_{11}$ & 1.13 &     $0.84 \pm 0.15$ & $0.98\pm 0.2$\\
o-l-C$_3$H$_2$ &     $2_{11}-1_{10}$ & 1.12 &         $ 0.84 \pm 0.15$ & $0.98\pm 0.2$\\                  
p-l-C$_3$H$_2$  &       $2_{02}-1_{01}$ & 0.41 &      $0.25 \pm 0.02 $  & $0.29 \pm 0.03 $ \\
p-c-C$_3$H$_2$  &       $2_{11}-2_{02}$ & 4.08 &   $2.5 \pm 0.4$ & $3 \pm 0.5$ \\

C$_4$H &             $4-3$          & 6.61 &             $5.35 \pm 0.02$ & $6.2 \pm 0.3$ \\
C$_4$H &          $5-4$ &        7.94 &            $5.21 \pm 0.05 $ &  $6.2 \pm 0.3$ \\
p-H$_2$CCO      &       $2_{02}-1_{01}$ & 4.19 &   $0.54 \pm 0.14$ & $0.62 \pm 0.15$  \\
o-H$_2$CCO      &       $2_{12}-1_{11}$  & 9.58 &  $1.9 \pm 0.2$ & $2.2 \pm 0.2$\\
o-H$_2$CCO      &       $2_{11}-1_{10}$ & 9.50 & $1.9 \pm 0.2$ & $2.2 \pm 0.2$\\
E-CH$_3$CHO     &       $2_{02}-1_{01}$ & 2.88 & $1.29 \pm 0.09$ & $1.5 \pm 0.1$ \\
A-CH$_3$CHO     &       $2_{02}-1_{01}$ & 2.80 & $1.35 \pm 0.09$ & $1.6 \pm 0.1$ \\
A-CH$_3$CN      &       $2_0-1_0$ & 0.64 & $ 0.19 \pm 0.02$ & $0.21 \pm 0.02$ \\
E-CH$_3$CN      &       $2_1-1_1$ & 0.75 & $ 0.18 \pm 0.02$ &  $0.21 \pm 0.02$ \\
HC$_3$N & $4-3$ & 1.20 & $0.2 \pm 0.04$ & $0.3\pm 0.1$ \\
HC$_3$N & $5-4$ & 1.38 & $0.4 \pm 0.1$ & $0.3 \pm 0.1$\\
HNCO & $2_{02}-1_{01}$ & 3.09 & $0.48 \pm 0.12$ & $0.56\pm 0.15$\\
SiO$^c$ & $1-0$          & 2.00$^c$ & $0.96 \pm 0.06$$^c$ & $1.1 \pm 0.08$$^c$ \\
\hline
\end{tabular}
\tablefoot{
   \tablefoottext{a}{Assuming an excitation temperature of 3.3~K}
      \tablefoottext{b}{Abundance relative to H$_2$ using N(H$_2$)=$8.6 \times 10^{20}$~cm$^{-2}$}
       \tablefoottext{c}{Assuming an excitation temperature of 6.5~K}
   }
\label{tab:col}
\end{table*}
}

\newcommand{\Tabelem}{
  \begin{table}
    \centering 
    \caption{Selected Abundance ratios toward BL Lac.}
\begin{tabular}{lc}
\hline \hline
Species & Abundance ratio \\
\hline
$^{12}$CS/$^{13}$CS & $104 \pm 25$\\
CS/C$^{34}$S & $27 \pm 5$ \\
CS/HCS$^+$ & $28 \pm 5$ \\
CS/H$_2$CS$^a$ & $21 \pm 6$\\
\hline
l-C$_3$H/C$_3$H$^+$ & $6.9 \pm 1.5$ \\
C$_4$H/l-C$_3$H & $2.65 \pm 0.2$ \\
C$_4$H/HC$_3$N & $ 26 \pm 6$ \\
\hline
H$_2$CO/H$_2$CCO$^b$ & $16 \pm 3$ \\
H$_2$CO/CH$_3$CHO$^b$ &$15 \pm 3$ \\
H$_2$CO/CH$_3$OH$^c$ & $>7.7$ \\
H$_2$CCO/CH$_3$OH$^c$ & $>0.5$ \\
CH$_3$CHO/CH$_3$OH$^c$ & $>0.5$ \\
\hline
\end{tabular}
\tablefoot{
 \tablefoottext{a}{Assuming an ortho to para ratio of 3 for H$_2$CS}
  \tablefoottext{b}{H$_2$CO data  from \citet{Liszt:1995}}
   \tablefoottext{c}{Upper limit on CH$_3$OH from \citet{Liszt:2008}}  
   }
\label{tab:elem}
\end{table}
}

\newcommand{\Tabab}{
  \begin{table*}[h!]
    \centering 
    \caption{Molecular abundances in diffuse and translucent clouds.}
\begin{tabular}{lccl}
\hline \hline
Molecule & Abundance & Range (dex) & Reference\\
 \hline
 OH      & $1.0 \times 10^{-7}$ & 0.1 & \cite{Weselak:2010} \\
  o-H$_2$O       & $2.7 \times 10^{-8}$ & 0.2 & \cite{Gerin:2019} \\
  OH$^+$ & $1.0 \times 10^{-8}$ & 0.5 & \cite{Jacob:2022} \\
  H$_2$O$^+$ & $1.0 \times 10^{-9}$ & 0.3 & \cite{Jacob:2022} \\
  H$_3$O$^+$ & $1.0 \times 10^{-9}$ & 0.3 & \cite{Lis:2014} \\
  CO &  $2.0 \times 10^{-6}$ & 1 & \cite{Liszt:2023} \\
  HCO & $1.0 \times 10^{-9}$ & 0.21 & \cite{Liszt:2025,Liszt:2014} \\
   HCO$^+$ & $3.0 \times 10^{-9}$ & 0.21 & \cite{Gerin:2019} \\
     HOC$^+$ & $4.6 \times 10^{-11}$ & 0.21 & \cite{Gerin:2019} \\
     H$_2$CO & $5.0 \times 10^{-9}$ & 0.21 & \cite{Gerin:2024} \\
       H$_2$CCO & $3.0 \times 10^{-10}$ &  & This work \\
       CH$_3$CHO & $3.1 \times 10^{-10}$ &  & This work \\
 \hline
  CH     & $3.6 \times 10^{-8}$ & 0.21 & \cite{Sheffer:2008} \\
   CH$^+$ & $5.0 \times 10^{-8}$ & 0.5 & \cite{Godard:2012} \\
  CH$_2$         & $1.3 \times 10^{-8}$ & 0.3 & \cite{Polehampton:2005} \\
  CH$_3$ & $2.0 \times 10^{-8}$ & 0.3 & \cite{Feuchtgruber:2000} \\
  C$_2$ & $5.0 \times10^{-8}$ & 0.3 & \cite{Fan:2024} \\
 C$_2$H & $4.4 \times 10^{-8}$ & 0.15 & \cite{Gerin:2019} \\
  C$_2$H$_2$ &  $1.5\times 10^{-8}$ & 0.3 & \cite{Feuchtgruber:2000} \\
  C$_3$ & $3.5 \times 10^{-9}$ & 0.3 & \cite{Fan:2024} \\
   l-C$_3$H & $2.1 \times 10^{-10}$ & 0.3 & \cite{Liszt:2018a} \\
     c-C$_3$H & $1.4 \times 10^{-10}$ & 0.3 & \cite{Liszt:2014} \\
     C$_3$H$^+$ & $5.0 \times 10^{-11}$ & 0.3 &  \cite{Gerin:2019}, This work \\
       c-C$_3$H$_2$ & $2.5 \times 10^{-9}$ & 0.3 & \cite{Liszt:2018a} \\
 l-C$_3$H$_2$ & $1.0 \times 10^{-10}$ & 0.3 & \cite{Liszt:2018a} \\
       C$_4$H & $6.2 \times 10^{-10}$ &  & This work \\
            C$_{60}$ & $8.5 \times 10^{-10}$ & 0.4 & \cite{Cordiner:2019,Berne:2013} \\
             C$_{60}$$^+$ & $5.0 \times 10^{-10}$ &  0.4 & \cite{Cordiner:2019,Berne:2013} \\
            \hline
     NH         &  $8.0 \times 10^{-9}$ & 0.2 & \cite{Persson:2012} \\
      o-NH$_2$  &  $4.0 \times 10^{-9}$ & 0.2 & \cite{Persson:2012,Persson:2016} \\
       o-NH$_3$         &  $2.0 \times 10^{-9}$ & 0.2 & \cite{Persson:2012,Persson:2016} \\
       N$_2$ & $3.0 \times 10^{-7}$ & 0.3 & \cite{Knauth:2004} \\
       CN & $3.0 \times 10^{-8}$ & 0.3 & \cite{Liszt:2001} \\
        HCN & $4.0 \times 10^{-9}$ & 0.3 & \cite{Liszt:2001,Liszt:2025} \\
         HNC & $1.1 \times 10^{-9}$ & 0.3 & \cite{Liszt:2001} \\
         CH$_3$CN & $4.5 \times 10^{-11}$ & 0.5 & \cite{Liszt:2018a},This work \\
         HC$_3$N & $2.2 \times 10^{-11}$ & 0.5 & \cite{Liszt:2018a} ,This work\\
         \hline
         SH & $1.1 \times 10^{-8}$ & 0.3 & \cite{Neufeld:2015}\\
 SH$^+$ & $1.4 \times 10^{-9}$ & 0.4 & \cite{Godard:2012}\\
  H$_2$S & $6.0 \times 10^{-9}$ & 0.3 & \cite{Neufeld:2015}\\
   CS & $3.0 \times 10^{-9}$ & 0.5 & \cite{Neufeld:2015,Lucas:2002}\\
   HCS$^+$ &  $1.0 \times 10^{-10}$ & 0.5 & \cite{Lucas:2002}\\
        SO & $1.0 \times 10^{-9}$ & 0.5 & \cite{Neufeld:2015,Lucas:2002}\\
           H$_2$CS & $1.5 \times 10^{-10}$ & 0.5 & This work\\
           CCS & $5.0 \times 10^{-11}$ & 0.5 & This work\\    
           \hline
        HF& $1.2 \times 10^{-8}$ & 0.14 & \cite{Gerin:2019,Sonnentrucker:2015}\\
 CF$^+$ & $1.7 \times 10^{-10}$ & 0.3 & \cite{Gerin:2019}\\
  HCl & $1.2 \times 10^{-9}$ & 0.3 & \cite{Monje:2013}\\
   HCl$^+$ & $1.0 \times 10^{-9}$ & 0.5 & \cite{Deluca:2012}\\
        H$_2$Cl$^+$ & $9.0 \times 10^{-10}$ & 0.5 & \cite{Deluca:2012,Neufeld:2015b}\\
           ArH$^+$ & $2.0 \times 10^{-10}$ & 0.5 & \cite{Schilke:2014}\\
          SiO & $1.0 \times 10^{-10}$ & 0.5 & \cite{Rybarczyk:2023}\\                    
\hline
\end{tabular}
    \label{tab:abun}
  \end{table*}
  }

\newcommand{\Figlines}{
  \begin{figure*}
    \centering 
    \includegraphics[width=0.85\linewidth]{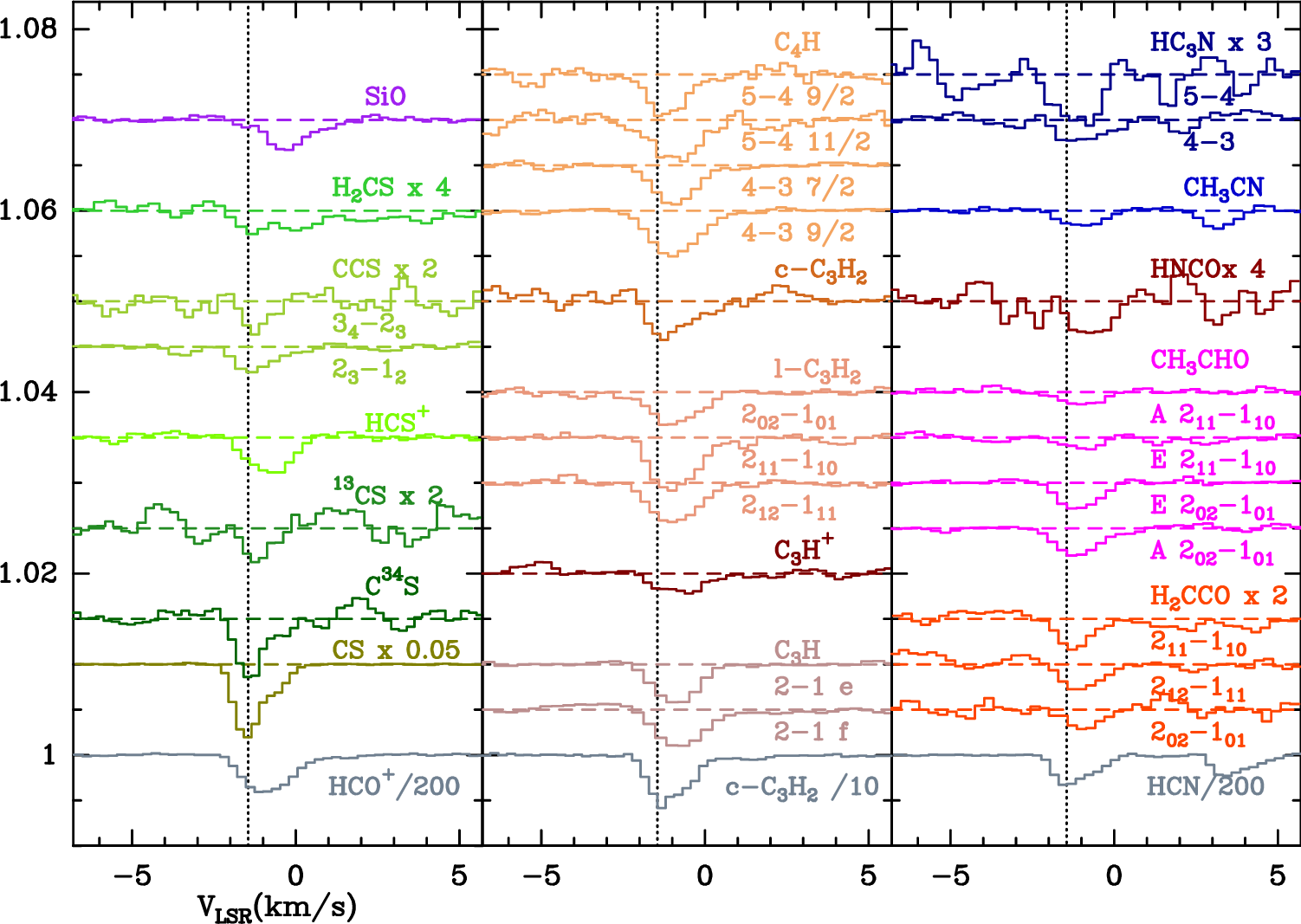}
    \caption{Absorption lines detected toward BL Lac in the Q band. All spectra are normalized by the continuum and  shifted vertically for clarity. The fainter lines are scaled as indicated. For comparison, we also show the absorption lines of HCO$^+$ ($J=1-0$) \citep{Lucas:1996}, HCN ($J=1-0$) \citep{Liszt:2001}, and c-C$_3$H$_2$ ($J_{K_1K_2} = 1_{1,0}-1_{0,1}$) \citep{Liszt:2012} in gray with their scaling factor. }
    \label{fig:lines}
  \end{figure*}}
  
\newcommand{\Figab}{
  \begin{figure*}
    \centering 
\includegraphics[width=0.85\linewidth]{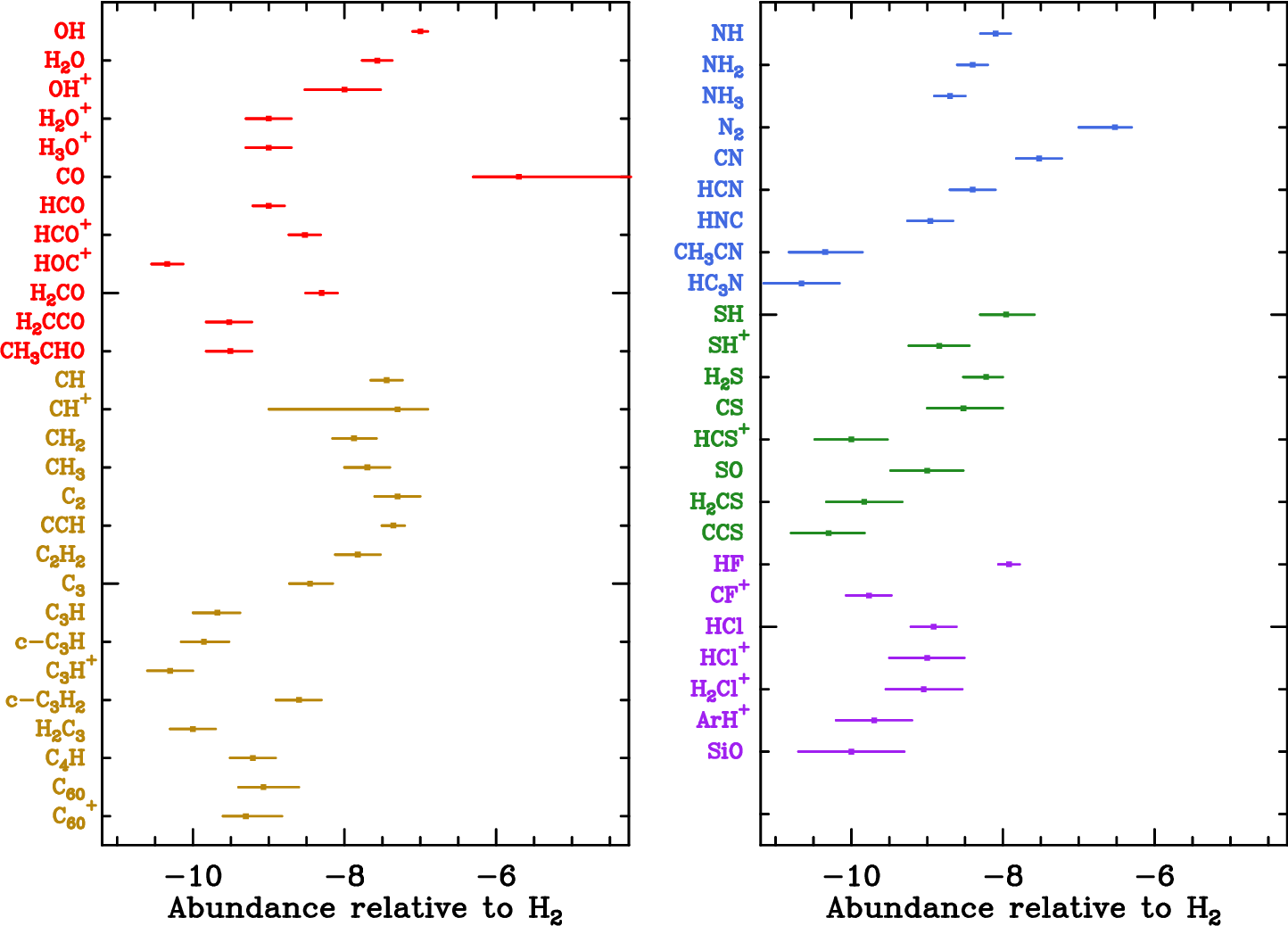}
    \caption{Summary of abundances of detected species in diffuse and translucent clouds as listed in Table \ref{tab:abun}.  }
    \label{fig:abun}
  \end{figure*}}

\newcommand{\Figcont}{
  \begin{figure*}[h!]
    \centering 
        \includegraphics[width=0.85\linewidth]{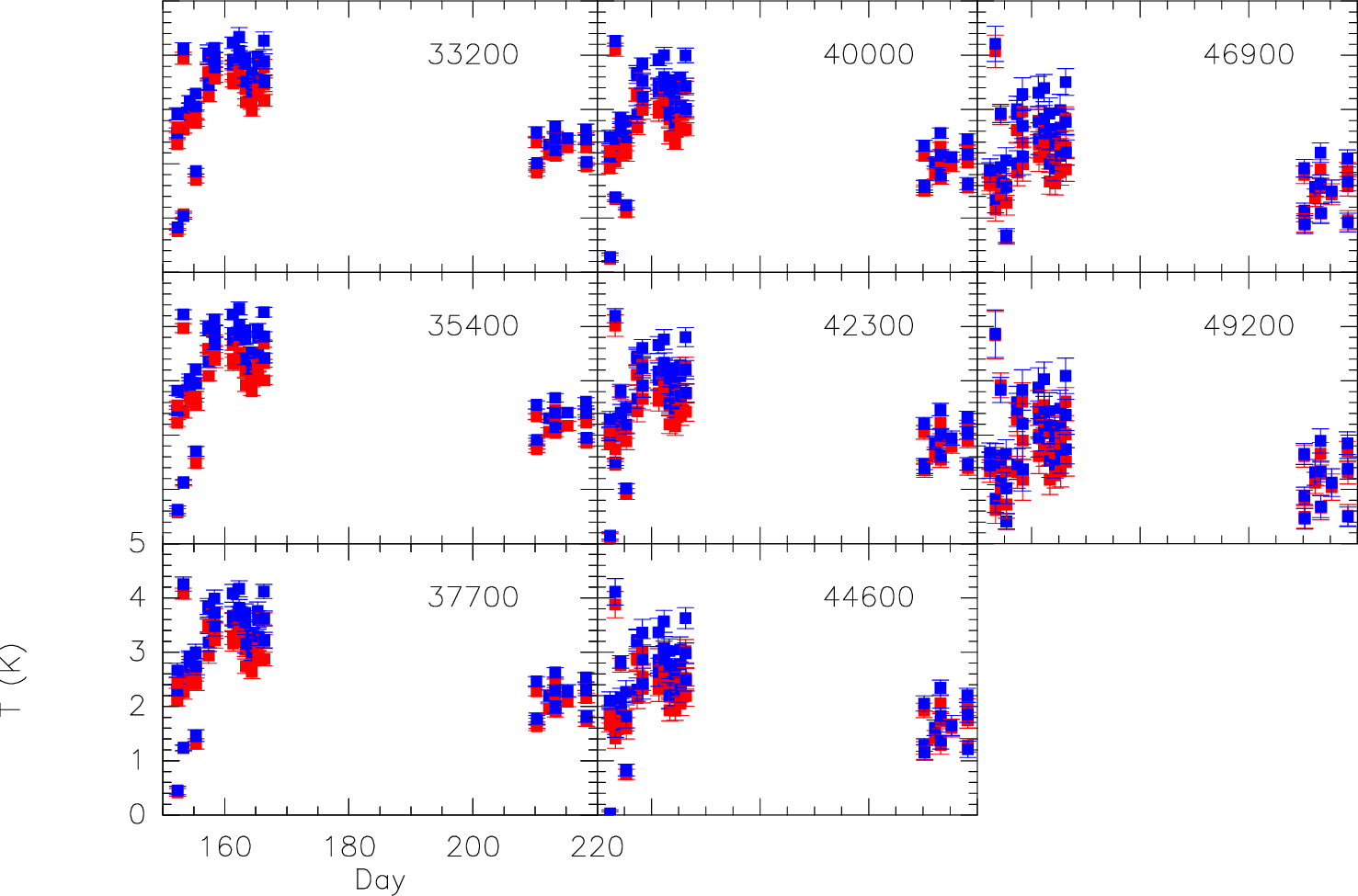}
    \caption{Variation of the continuum flux in H (red) and V (blue) polarization with time in 2024 for each spectral window. The central frequency of each spectral window in MHz is indicated in each panel. The gain
    varies from 3.7 Jy/K (at 32 GHz) to 4.7 Jy/K (at 48 GHz) (see \texttt{https://rt40m.oan.es/}).
   }
    \label{fig:cont}
  \end{figure*}}
  
  \newcommand{\Figpol}{
  \begin{figure*}[h!]
    \centering 
 \includegraphics[width=0.85\linewidth]{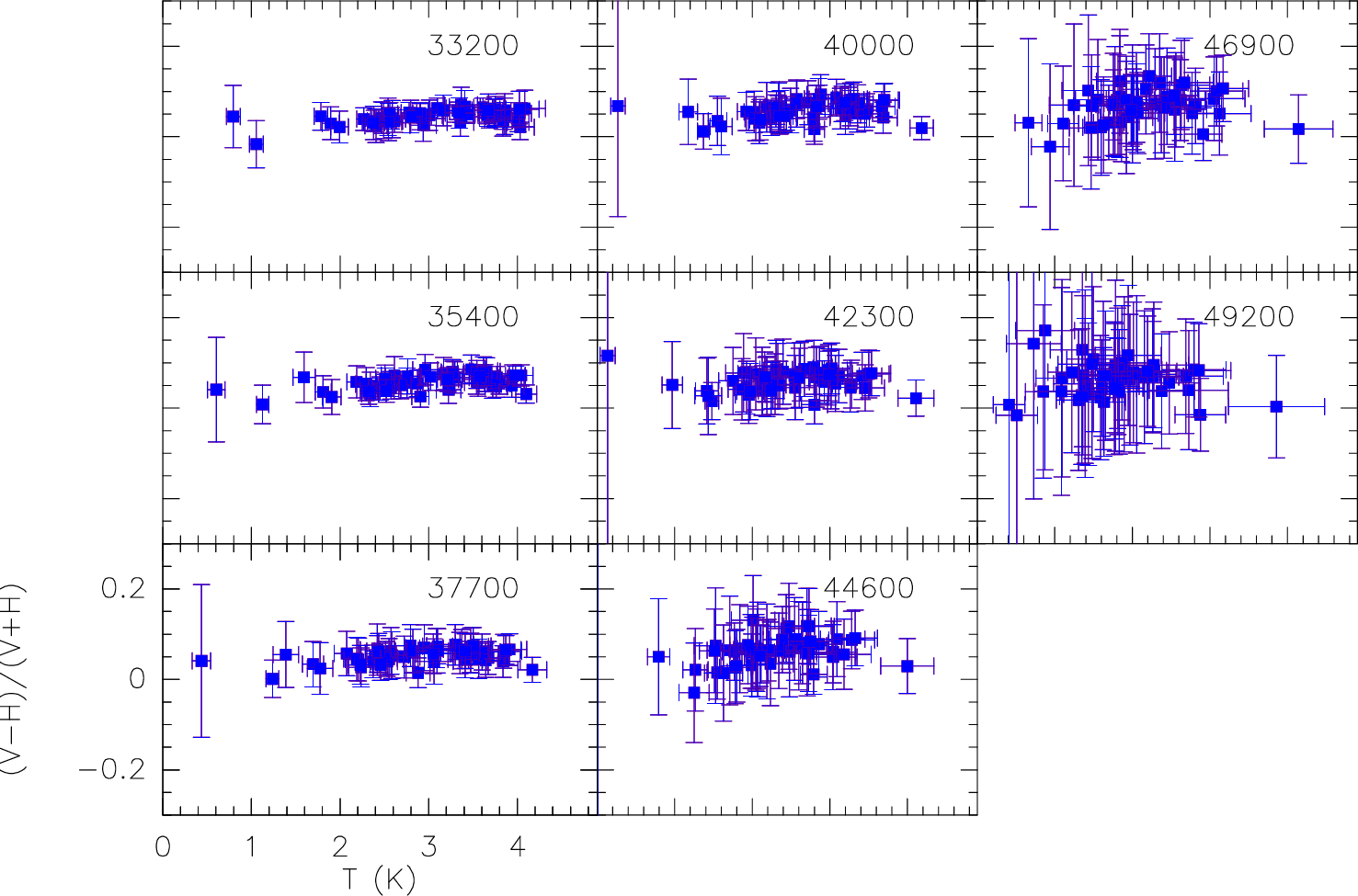}
    \caption{Variation of the polarization fraction $p$ with observed antenna temperature. The polarization fraction is defined as $p = \frac{T_V-T_H}{T_V+T_H}$ where $T_V$ and $T_H$ are the 
    continuum temperatures in the V and H polarizations, respectively.   }
    \label{fig:cont2}
  \end{figure*}}

    \newcommand{\Figspec}{
  \begin{figure}[h!]
    \centering 
        \includegraphics[width=0.9\linewidth]{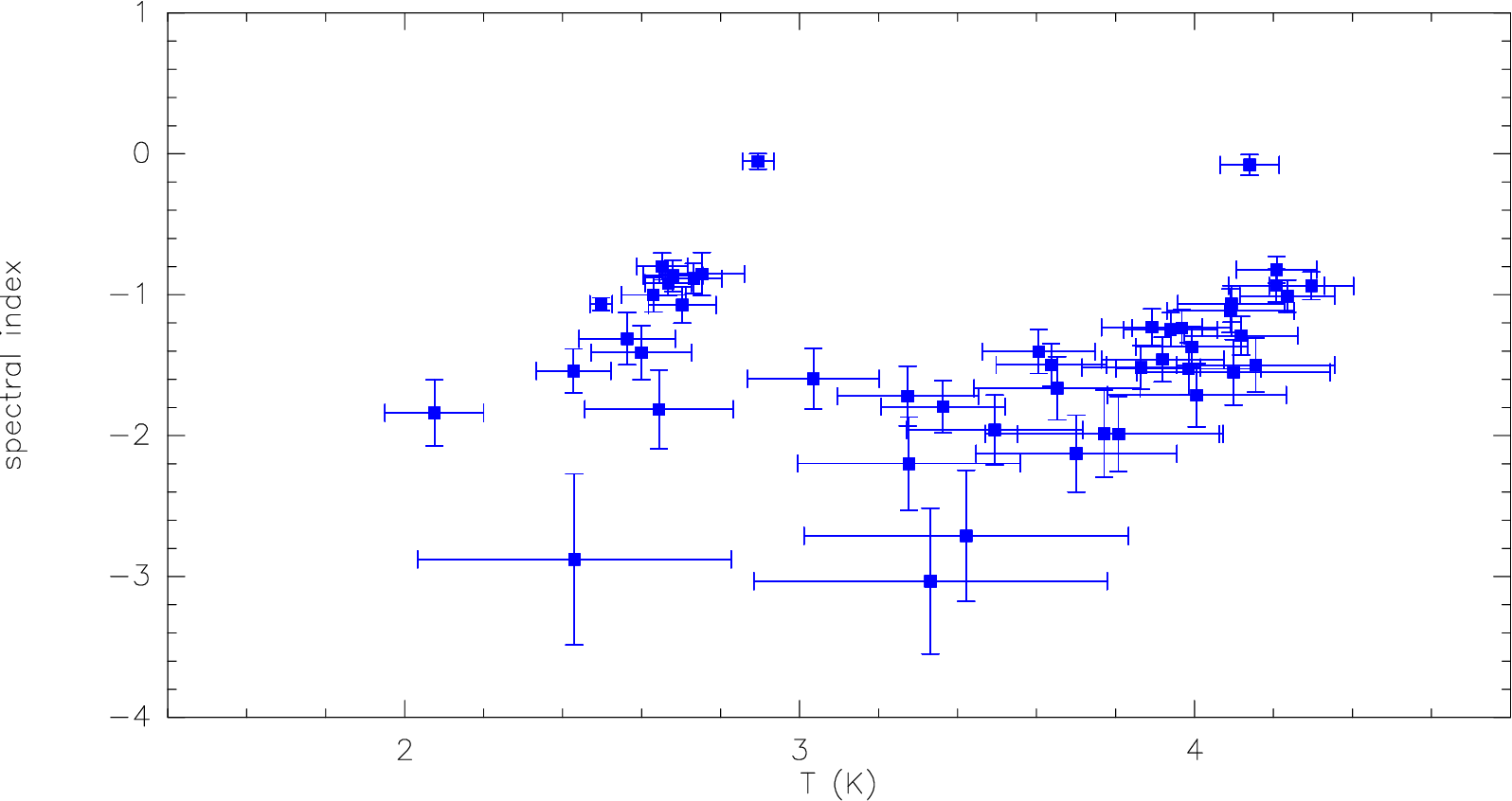}
    \caption{Variation of the spectral index of the continuum antenna temperature with the antenna temperature measured in the spectral window at the low end of the spectral range, 33.2 GHz.   }
    \label{fig:cont3}
  \end{figure}}

\begin{document} 

\title{Spectral survey of the diffuse gas toward BL Lac in the Q band}

\author{
  Maryvonne Gerin\inst{1} 
    \and Harvey Liszt\inst{2} 
  \and Bel\'en Tercero \inst{3,4} 
  \and Jos\'e Cernicharo\inst{5}
  }

\institute{LUX, Observatoire de Paris, PSL Research University, CNRS,
  Sorbonne Université, 75014 Paris, France.
   \and National Radio Astronomy Observatory, 520 Edgemont Road,
  Charlottesville, VA, 22903, USA. 
  \and  Observatorio Astron\'omico Nacional (OAN-IGN), Calle Alfonso XII, 3, 28014 Madrid, Spain.
\and Observatorio de Yebes (OY-IGN), Cerro de la Palera s/n, 19141 Yebes, Guadalajara, Spain.
   \and  Instituto de Física Fundamental, CSIC, Calle Serrano 123, 28006 Madrid, Spain. 
}

\date{Received 2025; accepted yy   2025}

\abstract
{The chemical composition of diffuse interstellar clouds is not fully established. They host an active chemistry despite their relatively low density and the ubiquitous presence of far-UV radiation. 
}
{To further explore the chemical composition of diffuse clouds, we performed a spectral scan toward the bright radio source BL Lac in the Q band (from 32 to 50 GHz) using the Yebes 40m telescope.}
{Yebes observations were performed interleaving frequency switching and position switching integrations toward BL Lac, using a spectral resolution of 38 kHz. The data were reduced with the CLASS software. }
{We achieved an unprecedented sensitivity on the continuum of 0.02 - 0.07 \%, allowing for the detection of very faint absorption features. We confirm previous detections of HCS$^+$, C$_3$H, C$_3$H$^+$, CH$_3$CN, and HC$_3$N in diffuse clouds and report new detections of CCS, C$_4$H, CH$_3$CHO, H$_2$CCO, HNCO, and H$_2$CS along the line of sight to BL Lac, with abundances relative to H$_2$ from a few 10$^{-11}$ to a few 10$^{-10}$. We compiled molecular detections toward diffuse clouds to obtain the chemical inventory of a typical diffuse interstellar cloud.  }
{The chemical inventory of diffuse interstellar clouds includes complex organic species with up to four heavy atoms. These species are efficiently formed in the diffuse interstellar gas and reach abundances similar to those measured in dense photodissociation regions, pointing to similar gas-phase chemical processes. }


\maketitle{} 
\begin{nolinenumbers}
\section{Introduction}
%\LEt{A\&A uses the past tense to describe specific methods used in a paper, and the present tense to describe general methods and the findings of recent papers. Furthermore, the use of the present tense as opposed to the future is advised for generalities and when authors discuss what can be found within their own work. For more details, see Sect. 6 of the language guide https://www.aanda.org/for-authors/language-editing/6-verb-tenses. Please review my edits to ensure this was carried out appropriately throughout your paper and make any further necessary edits.***}

Diffuse interstellar clouds seen on sight lines at $A_V \leq 1$ mag had long been thought to be chemically simpler than molecular clouds because they are fully bathed by \rev{Far Ultra Violet (FUV)} 
%\LEt{Please remember to spell out acronyms upon first appearance in the main text. Thereafter the acronym should be used unless it’s at the beginning of a sentence. Please make the necessary changes throughout. ***} 
radiation and have moderate densities up to a few hundreds cm$^{-3}$ \citep{Snow:2006}. In diffuse clouds, dust grains are bare, and no ice mantle can develop, as shown by the threshold for the detection of water ice at $A_V \sim 3$ mag \citep{Boogert:2015}. However, this simple view, based on UV \rev{and}
%\LEt{We reserve the use of slashes to denote ratios and instrument or wavelength pairings and for use in equations. The use of "and/or" is also acceptable. Kindly check the edits here and throughout and rephrase where needed.***} 
optical spectroscopy, was upset by radio-frequency absorption line spectroscopy toward compact extragalactic continuum sources. In addition to the diatomic species OH, CH, CH$^+$, CO, and CN detectable at UV and optical wavelengths, c-C$_3$H$_2$ \citep{Cox:1988}, NH$_3$, and H$_2$CO \cite{Nash:1990}, HCO$^+$, HNC, HCN, and C$_2$H \citep{Lucas:1994, Lucas:1996, Lucas:2000} were found along sight lines with $A_v = 1$~mag far from the Galactic plane. Additional surveys have shown that SO, CS, and CF$^+$, and the polyatomic species HOC$^+$, 
c-C$_3$H, l-C$_3$H, l-C$_3$H$_2$, C$_3$H$^+$, and CH$_3$CN \citep{Liszt:2014,Liszt:2018a,Gerin:2019} are detectable with abundances relative to H$_2$ ranging from 10$^{-10}$ to 10$^{-8}$, reaching levels that are surprisingly close to those of the TMC-1 dark cloud, as cataloged for instance by \cite{Gratier:2016}. The QUIJOTE\footnote{Q-band Ultrasensitive Inspection Journey to the Obscure TMC-1 Environment} spectral survey of the TMC-1 dark cloud allowed for a comparison of the relative abundances of cumulene species, such as l-C$_3$H$_2$, l-C$_4$H$_2$, and l-C$_5$H$_2$ \citep{Cabezas:2021}, and showed that the first members of the series reach abundances of a few 10$^{-10}$ with respect to H$_2$. It is interesting to note that some ions are more abundant in the diffuse gas than in TMC-1, especially the important precursor of the carbon chains, C$_3$H$^+$, which is about one order of magnitude more abundant in the diffuse gas than in TMC-1 \citep{Cernicharo:2022}. In addition, the HIFI instrument on board the {\it Herschel} space mission and the GREAT instrument carried on the SOFIA airplane led to the detection of neutral and ionized hydrides in diffuse interstellar clouds, including the surprising detection of reactive ions such as OH$^+$ and ArH$^+$ \citep[for a review see][]{Gerin:2016}.

Numerical simulations of the formation of molecular clouds from atomic gas show that the physical structure of diffuse clouds is related to the transient multiphase structure of the interstellar medium as a whole, with regions of partially-molecular cool gas (the cold neutral medium or CNM) embedded in the warm neutral medium (WNM) \citep[e.g.,][]{Valdivia:2017,Godard:2023}. These simulations reproduce the statistics of the H$_2$ column densities observed in UV absorption against background massive stars but do not succeed with the current chemical networks in reproducing the observed abundances of molecular species such as HCO$^+$, c-C$_3$H$_2$, and CH$_3$CN \citep{Liszt:2018a}. The case of HCO$^+$, which is the immediate ancestor of CO \citep{Visser:2009}, is particularly important: this ion is widespread and easily detected in absorption toward background quasars \citep{Lucas:1996,Liszt:2023}. These failures show that the models are not yet fully adequate. However, it has been difficult to identify the missing physical and chemical processes because systematic inventories of the composition of template diffuse clouds are not available in the same way as those established for dark clouds or hot cores. The lack of a more complete inventory prohibits searching for relationships between molecules and establishing families of species that could be used for identifying the dominant physical and chemical processes.

The availability of a sensitive broadband 32-50 GHz receiver at the Yebes 40m telescope offers the opportunity
to detect different classes of species than those accessible at millimeter frequencies.
The success of the QUIJOTE survey performed with this system in detecting new species, in particular polyatomic molecules, 
in TMC-1 and other dark clouds \citep[e.g.,][]{Cernicharo:2021a, Cernicharo:2021b, Agundez:2023}
shows that many species can reach abundances relative to H$_2$ higher than $10^{-10}$. Given the low densities of diffuse clouds, the lowest-energy rotational levels are
the most populated, which implies that the most favorable frequency domain shifts toward frequencies below 50~GHz for species with
three or more heavy atoms. In this paper, we present the outcome of a deep absorption survey toward the BL Lac (also known as B2200+420) radio source in the Q band. Section \ref{sec:obs} presents the observations and data reduction process; Sect. \ref{sec:results} presents the detected spectral features, the derived column densities, and the relative abundances. The new
detections are put in the context of molecular abundances in diffuse clouds in Sect. \ref{sec:discussion}, and 
Sect. \ref{sec:conclusion} presents the final conclusions.

\section{Observations and data reduction}
\label{sec:obs}

The sight line toward BL Lac is known to exhibit intervening absorption by foreground diffuse clouds, with a reddening of E(B-V) = 0.32 mag corresponding to one magnitude of visual extinction, a total hydrogen column density of about $\rm{N(H)}=\rm{N(HI)+2N(H_2)} = 2.7 \times 10^{21}$~cm$^{-2}$, and a relatively high fraction of hydrogen in molecular form $f(\rm{H_2}) = \frac{\rm{2N(H_2)} } {\rm{N(H)}} $ of 0.66 \citep{Liszt:2012b}. At a Galactic latitude of $-10.4^\circ$, the absorbing gas is located within 1 kpc \citep{Green:2019,Lallement:2022}, with a likely distance of $\sim 500$~pc.
HI absorption is detected across the velocity range between $-10$ and $10$ \kms, while molecular absorption is more localized between $-3$ and $1$ \kms in the blue wing of the HI absorption. CO emission and absorption lines are quite strong along this sight line, in agreement with the high fraction of gas in molecular form. 
\Tabobs

We used the Yebes 40m telescope to perform an absorption survey toward the BL Lac radio source in April, May, and June 2024 under the project code 24A023 as summarized in Table \ref{tab:obs}.
The observations were performed in frequency switching (FSW) mode, using a frequency throw of 10.52~MHz.
In addition, every $\sim$ 1.25 hours, we checked the pointing and performed position switching (PSW) observations with a reference position located at equatorial coordinates (J2000) \rev{Right Ascension (RA)}%\LEt{Please define the acronym. ***} 
= $22^h02^m36^s$ and \rev{Declination (DEC)} %\LEt{Please define the acronym. ***}
 = $+42^{\circ}15'30''$, a molecular line emission-free position nearby the target source (see \citealt{Liszt:2020}), to measure the continuum flux and follow the variations of the telescope efficiency with the ambient temperature and telescope elevation. 

The telescope pointing was checked using BL Lac and the nearby HII region DR21 as pointing sources. BL Lac was flaring during the time of these observations with a peak temperature of $\sim 4~$K corresponding to $\sim $ 14\,Jy at 32 GHz, which allowed us to check the pointing accuracy on the source itself with high confidence.
To avoid spurious signals and %\LEt{Please verify that the intended meaning has not changed.***} 
sideband ambiguities in the detected line frequencies and better cover the spectral range, we used two frequency tunings, with a separation of 900\,MHz in the local oscillator frequencies. 
The Q-band receiver allowed us to measure the spectral range from 32~GHz to 50~GHz with a spectral resolution of 38~kHz (equivalent to 0.356 \kms at 32 GHz), observing the H and V polarizations simultaneously. 
The backend is composed of a series of eight \rev{Fast Fourier Transform spectrometer (FFT)} %\LEt{Please define the acronym. ***}
 elements per polarization, for a total of sixteen spectral windows.
The system and telescope performances have been described in \citet{Tercero:2021}.
The data were reduced with the CLASS software, which is part of the GILDAS package\footnote{\texttt{http://www.iram.fr/IRAMFR/GILDAS}}. For each day of observations, we analyzed the data as follows. For each frequency tuning and each of the 16 spectral windows, the spikes leading to spurious signals were flagged and the spectra interpolated in the small frequency intervals of these spikes. The PSW data and FSW data were separately averaged to obtain a PSW and a FSW spectrum for each spectral window. The FSW data were folded at this step. We removed a continuum offset (flat baseline) and a sine wave function associated with a known baseline ripple in this observing mode.
To produce continuum-divided spectra, we first smoothed the PSW spectra in each spectral window to a coarse spectral resolution of 380\,MHz to obtain a noise-free continuum spectrum. We then added the continuum to the FSW spectra and co-added the full spectral resolution FSW and PSW spectra to obtain the source spectrum. We divided this spectrum by the noise-free, smoothed PSW continuum spectrum to obtain the absorption spectrum normalized to the continuum. We performed these operations each day and for each polarization because the continuum flux of BL Lac is variable and presents significant polarization. Information on the BL Lac continuum flux during the observations is provided in Appendix \ref{app:cont}.

In the final step, we averaged the data gathered from the different days and from the two polarizations, and combined the two frequency tunings to obtain the final continuum-divided spectrum that was used for line detection. The resulting noise level on the normalized spectrum is $3 \times 10^{-4}$ with the lowest values below 40 GHz where the sky transmission is the best and increases to $1 \times 10^{-3}$ near the CS(1-0) line at 49 GHz. 

Given the high brightness of BL Lac in the Q band at the Yebes Observatory, with typical zenith opacities during our observations ranging from $\sim 0.05$ at 32 GHz to $\sim 0.3$ at 49 GHz, this procedure provides accurate relative continuum measurements. The good agreement of the column densities derived in this work (see Sect. \ref{sec:results}) for species detected at millimeter frequencies with PdBI or NOEMA such as CS and c-C$_3$H$_2$ \citep{Lucas:2000,Lucas:2002}, as well as the agreement of integrated opacities for the spectral lines previously observed with
\rev{ the Very Large Array (VLA)} %\LEt{Please consider defining.***} 
\citep{Liszt:2018a} shows that this procedure provides reliable continuum normalized spectra.
\rev{Table \ref{tab:obs} presents a summary of the observations discussed in this paper.}
%\LEt{As per A\&A's guidelines to avoid single-sentence paragraphs, please consider merging this with the paragraph above.***}

\Figlines

\section{Results}
\label{sec:results}

\subsection{Line identification}

\Tabcol

The data were inspected for spectral features using the CDMS \citep{Muller:2005,Endres:2016}, JPL \citep{Pickett:1998}, and MADEX %\LEt{Please consider defining the acronyms.***} 
\citep{Cernicharo:2012} line catalogs as references for the molecular line frequencies. Figure \ref{fig:lines} shows the detected absorption lines together with earlier detections of HCO$^+$ (J=1-0)\citep{Lucas:1996}, HCN (J=1-0) \citep{Liszt:2001}, and c-C$_3$H$_2$ ($1_{1,0}-1_{0,1}$) \citep{Liszt:2012} for comparison. Table \ref{tab:lines} reports the information on the detected lines in the Q band. As two velocity components are identified along this sight line in some species, the peak opacities and integrated line opacities were determined by fitting either one or two Gaussian profiles to the detected signals, according to the achieved velocity resolution. We checked the quality of the fit by comparing the fitted values with the integrated opacity computed in the velocity range where HCO$^+$ absorption is detected from $-3$ to $0$ \kms. 

The sensitive Q-band survey allows the detection of new spectral lines of already known species such as CS and its isotopologs C$^{34}$S and $^{13}$CS, SiO, the carbon chains and cycles with three and four carbon atoms C$_3$H, C$_3$H$^+$, C$_4$H, c-C$_3$H$_2$, and l-C$_3$H$_2$. These new data confirm the previous detection of CH$_3$CN toward BL Lac but with an improved signal-to-noise ratio and add a new detection of HC$_3$N previously found toward the higher extinction sight line of 3C111 \citep{Liszt:2018a}. The surprise comes from the detection of two organic species thought to be associated with star-forming cores. We find three lines from ketene (H$_2$CCO) corresponding to its ortho and para spin symmetry states, and six lines from acetaldehyde (CH$_3$CHO) associated with its A and E states. A tentative feature can be assigned to a low-energy transition of HNCO and another one to the ground state transition of H$_2$CS. Two transitions of the carbon and %\LEt{Please verify that the intended meaning has not changed.***} 
sulfur chain CCS, known to be abundant in TMC-1, are detected. The new observations also show absorption from HCS$^+$, which has only previously been detected toward 3C111.

We searched for features from related species, which \rev{have} transitions in the covered spectral range, CCO, CCN, CH$_3$CCH, C$_3$N, C$_4$H$^{-}$, and C$_6$H, but find no convincing signal down to the noise level. The achieved noise levels are reported in Table \ref{tab:upper}. We also provide upper limits on the relative abundance for these species assuming an excitation temperature of 5~K as a conservative value for the moderate densities encountered along the sight line to BL Lac.

The line profiles are consistent with the known centroid velocity of previously detected absorption lines toward BL Lac near $V_{LSR}=-1$~\kms, but different species show different line profiles. Two velocity components are clearly present in the CS and c-C$_3$H$_2$ data, at $V_{LSR} = -1.5$ and $-0.6$~\kms, with the stronger absorption at the more negative velocity, $-1.5$~\kms. The profiles of the HCS$^+$ and C$_3$H$^+$ molecular ions show stronger absorption in the $-0.6$~\kms velocity component, leading to slightly redshifted profiles compared to those of CS. Because of the slightly coarser velocity resolution, the profiles of the lines with frequencies lower than about 35~GHz show only one velocity component. The mean full width at half maximum (FWHM) when fitting a single Gaussian component is $1.3$~\kms. SiO is a clear outlier with a slightly broader profile than the mean, FWHM=1.4~\kms, and a different centroid velocity, $V_{LSR}=-0.2$~\kms, as compared to the centroid velocity for other species between $-1.5$ and $-0.6$~\kms. This behavior is consistent with the SiO formation pathway involving low-velocity shocks \citep{Rybarczyk:2023}.

 Such differences in the line profiles were also seen toward other sight lines and indicate the presence of density and velocity gradients along the line of sight \citep[][e.g., for the sight line to NRAO150 (B0355+508)]{Gerin:2024}. 
The abundance variations are related to the different chemical pathways responsible for the formation and destruction of the different species. Species such as CS have an enhanced abundance in the regions with the highest density along the line of sight, which have a smaller velocity dispersion compared to the total velocity dispersion of the molecular gas traced by CCH or HCO$^+$ \citep{Liszt:2025}. Carbon chains and cycles seem to occupy the full volume and exhibit a velocity dispersion similar to that of HCO$^+$ or CCH. Some molecular ions are more abundant in the lower density regions where the electron fraction is the highest, especially those reacting with H$_2$ such as OH$^+$ and CH$^+$ \citep{Gerin:2016}. The observed profile of C$_3$H$^+$, an ion reacting with H$_2$ to form C$_3$H$_2^+$ and C$_3$H$_3^+$ \citep{Guzman:2015}, follows this trend as the velocity component at $-0.6$ \kms is more prominent than that at $-1.5$ \kms.
The same behavior is seen for HCS$^+$. It could indicate a difference in density and ionization fraction between these two velocity components, since the abundance of this molecular ion is expected to be higher in lower-density gas with a high electron fraction \citep{Lucas:2002}.

\subsection{Molecular abundances}

We report in Table \ref{tab:col} the column densities of the detected species. Except for SiO, we use an excitation temperature of 3.3~K for deriving column densities from the integrated line opacities. As discussed in \citet{Liszt:2018a}, at the relatively low frequencies of the Q band, collisional excitation by electrons and neutrals is effective especially for relatively long chains such as C$_4$H or HC$_3$N. The chosen value of the excitation temperature is derived from the two well-detected doublets of C$_4$H, $n=4-3$ near 38~GHz and $n=5-4$ near 47~GHz. The CS excitation temperature derived by comparing the present observations of the $J=1-0$ transition with earlier measurements of the $J=2-1$ line \citep{Lucas:2002} is very similar to that of C$_4$H.
SiO presents a higher excitation temperature of $T_{ex} = 6.5_{-2}^{+6}$~K, when comparing our measurements of the $J=1-0$ line with observations of the $J=2-1$ line near 86~GHz with the \rev{Northern Extended Millimeter Array (NOEMA)} % \LEt{Please consider defining the acronym.***} 
\citep{Rybarczyk:2023}. 
For species where two velocity components are detected, we derived the total column density by combining the integrated opacities of these two components. When several transitions were available, we computed the mean column density as a weighted average of the column densities derived from each transition and used this value to compute the species abundance relative to H$_2$. 
We derived abundances relative to H$_2$ using $\rm{N(H_2)} = 8.6 \times 10^{20}$ cm$^{-2}$, the column density estimated from the HCO$^+$ absorption, along with the mean abundance for HCO$^+$, $3\times 10^{-9}$ \citep{Gerin:2019}.
%\LEt{Please verify that the intended meaning has not changed.***}  

Table \ref{tab:elem} presents a selection of abundance ratios. The relative abundances of the CS isotopologs follow those of the carbon and sulfur isotopes, within the accuracy of the present measurements; namely, $^{12}$CS/$^{13}$CS $\sim$ $^{12}$C/$^{13}$C and C$^{32}$S/C$^{34}$S $\sim$ $^{32}$S/$^{34}$S. 
HCS$^+$ has a fairly high column density, similar to that of C$^{34}$S. A similar value of the CS/HCS$^+$ abundance ratio was previously found along the line of sight to 3C111 \citep{Lucas:2002}. As discussed in \cite{Lucas:2002}, when CS/HCS$^+$ is approximately $20$, the dissociative recombinations of HCS$^+$ ions with the abundant electrons in the diffuse gas are efficient enough to produce the observed CS abundance. 
Using reaction rates in the KIDA database\footnote{\texttt{https://kida.astrochem-tools.org/} \cite{Wakelam:2012}}, the free-space photodissociation rate of CS is 10$^{-9}$ s$^{-1}$ and the rate of the dissociative recombination of HCS$^+$ with electrons leading to CS is $1.80 \times 10^{-7} (T/300)^{-0.57}$ cm$^3$s$^{-1}$. By balancing the CS photodissociation by the formation through the HCS$^+$ dissociative recombination, we get a relationship between the CS and HCS$^+$ abundances X(CS) and X(HCS$^+$):
$$X(\rm{CS}) = 14 \frac{x(e)}{10^{-4}}\frac{n(\rm{H_2})}{100 \, \rm{cm^{-3}}}\frac{0.5}{f(H_2)}\left(\frac{T}{30}\right)^{-0.57} X(\rm{HCS^+}).$$

The observed abundance ratio between CS and HCS$^+$ of $\sim 28$ indicates that HCS$^+$ can be considered a precursor of CS, as HCO$^+$ is a precursor of CO in diffuse and translucent clouds. This raises questions on the formation pathways of these precursor molecular ions.

\Tabelem

The species with the largest number of carbon or nitrogen atoms, C$_4$H and HC$_3$N, show a contrasting behavior. C$_4$H is more abundant than C$_3$H either in its linear (this work) or cyclic \citep{Liszt:2014} forms, a pattern seen for carbon chains in dark clouds such as TMC-1 
\citep[][e.g.,]{Oyama:2020}. While the CCH abundance derived by \cite{Lucas:2000} is X(CCH) =$3.6 \times 10^{-8}$, C$_4$H is about 50 times less abundant than CCH, indicating a steeper decrease in the carbon chain abundance with respect to the total number of carbon atoms compared with dark clouds where C$_2$H/C$_4$H is approximately $10$ \citep{Oyama:2020, Liszt:2018a}. 
HC$_3$N presents a deficit with respect to carbon chains toward BL Lac, consistent with the first detection in the translucent cloud along the line of sight to 3C111 \citep{Liszt:2018a}, as we measure C$_4$H/HC$_3$N $\sim 26\pm 10$, while the same ratio is close to $2$ in TMC-1. Similar column densities for C$_4$H and HC$_3$N are clearly excluded from our data, as this would imply stronger absorption features from HC$_3$N, which would have been clearly detected at the sensitivity reached in the survey.

The most remarkable detections are those of the organic species ketene (H$_2$CCO) and acetaldehyde (CH$_3$CHO). These molecules reach comparable column densities of $\sim 2.5 \times 10^{11}$~cm$^{-2}$ along the line of sight to BL Lac. They are about 15 times less abundant than formaldehyde. Ketene and acetaldehyde have previously been detected in dense photodissociation regions such as the Orion Bar and the Horsehead Nebula \citep{Guzman:2014,Cuadrado:2017}, indicating that they can resist FUV radiation. It is interesting to note that ketene and acetaldehyde reach similar abundances  in the Orion Bar and the Horsehead Nebula, about 10$^{-10}$ relative to H$_2$, a factor of 10 lower than formaldehyde along the same lines of sight. Their abundances relative to H$_2$ along the sight line to BL Lac also reach the same level, at a few times 10$^{-10}$, indicating that ketene and acetaldehyde are formed at the same efficiency in all these UV-illuminated environments. In TMC-1, ketene is about four times more abundant than acetaldehyde reaching an abundance of $1.5 \times 10^{-9}$ with respect to H$_2$ \citep{Cernicharo:2020}. The abundance of acetaldehyde appears remarkably uniform in these different environments, remaining at a level of $\sim 3 \times 10^{-10}$ relative to H$_2$. With the low extinction along the sight line to BL Lac, dust grains are not expected to host ices; hence, the most likely formation pathway must involve gas-phase chemistry. The conditions for forming ketene are more favorable in the denser and more shielded environment of TMC-1.

\section{Discussion}
\label{sec:discussion}

\Figab

To put the sight line to BL Lac into the context of diffuse and translucent clouds, we compile the abundances of molecular species detected by the absorption lines in diffuse and translucent clouds in Table \ref{tab:abun}. We include neutral and ionized fullerene C$_{60}$ and C$_{60}^+$, as they are representative of large carbonaceous species. Both were spectroscopically identified through their absorption bands for C$_{60}^+$ \citep{Cordiner:2019}, and through specific emission bands in high-resolution mid-infrared spectra of reflection nebulae for C$_{60}$ and C$_{60}^+$ \citep{Berne:2013}. 
Figure \ref{fig:abun} presents a graphical display of the same information. As the detection methods and source samples are different for the different species, high-mass stars, radio sources, and compact HII regions, this compilation does not represent the abundance of a “typical” sight line through diffuse gas. It is meant to summarize the variety of chemical species already identified and the order of magnitude of their abundances. Some species, such as CH, OH, and CH$^+$, can be measured using different techniques accessible both at visible and radio and/or submillimeter wavelengths, which allows for checking the consistency of the abundance measurements for the different samples of background sources.%\LEt{Please verify that the intended meaning has not changed.***} 
The range of abundance variations is also indicated as a logarithmic interval. Abundance variations are ubiquitous in diffuse and translucent clouds from sight line to sight line or among velocity components along the same sight line, as the abundances are related to the detailed local physical conditions. For instance, the sight line toward NRAO 150 hosts five velocity components of similar total H$_2$ column densities, two of which have stronger features in CS and H$_2$CO associated with denser gas than the three other components \citep{Gerin:2024}.
%\LEt{Please verify that the intended meaning has not changed.***}

The new detections achieved toward BL Lac nicely fit into previous detections and confirm that relatively complex species such as C$_4$H or acetaldehyde reach abundances relative to H$_2$ higher than 10$^{-10}$ even in the low-density, low-extinction diffuse and translucent clouds. These abundances are comparable to the abundance of neutral and ionized fullerene C$_{60}$ and C$_{60}^+$ \citep{Cordiner:2019,Berne:2013}. It is therefore tempting to bridge the gap between the small gas-phase species and the larger species known to also be present in diffuse and translucent clouds, such as polycyclic aromatic hydrocarbons (PAHs), fullerenes, and the carriers of the diffuse interstellar bands (DIBs). Recent progress on the spatial distribution of selected DIBs and their relation to the 3D structure of the interstellar medium \citep{Lallement:2024} now provides detailed information on the location of the DIB carriers with respect to the diffuse clouds and their association with either atomic or molecular gas. Some DIBs seem to be associated with diffuse molecular gas, while others better follow the lower-density atomic gas \citep{Lallement:2024}. Because the extinctions are too low for ice mantles to be present, molecule formation on dust surfaces is excluded in diffuse clouds. However, the similar abundances between complex species and fullerene suggest a chemical relation. Current chemical models do not include the chemical coupling between the small gas-phase species, the  PAHs, and related fullerenes, lacking detailed information on the rates and products of the potential chemical processes. Theoretical calculations of carbon atoms accreting on PAHs show that the resulting complexes after several accretion steps may host side chains that could be released in fragments of various sizes in the gas phase \citep{Omont:2025}. The losses of C$_2$H and C$_2$H$_2$ are calculated to be important channels and could represent a significant fraction of the total formation rate of these two molecules in the diffuse molecular gas %\LEt{Please define the acronym. ***}. 
C$_2$H and C$_2$H$_2$ can further be processed or react with ions and neutrals to build heavier carbon species. Accretion of oxygen atoms is also possible, although less favored than the accretion of carbon atoms. Further studies on the reactivity of PAHs and fullerenes in their neutral or ionized forms with carbon and oxygen are needed for a more quantitative assessment of the role of these species in the production of gas-phase small molecules.

\section{Summary and conclusions}
\label{sec:conclusion}

We presented a deep spectral survey in the Q Band toward the bright radio source BL Lac. Thanks to the source flaring at the time of the observations, we achieved an unprecedented sensitivity on the continuum of 0.02 - 0.07 \% in 38~kHz channels. At this sensitivity, we reported new detections in the diffuse medium along the line of sight to BL Lac of CCS, C$_4$H, CH$_3$CHO, H$_2$CCO, and H$_2$CS and confirmed previous detections of C$_3$H$^+$, HCS$^+$, CH$_3$CN, and HC$_3$N.

The abundances relative to H$_2$ of the organics ketene and acetaldehyde are comparable to those observed in such dense photodissociation regions as the Horsehead Nebula or the Orion Bar, suggesting that similar gas-phase chemical processes could be at work in these UV-illuminated environments. %\LEt{Single-sentence paragraph. ***}

The decrease in the abundance of carbon chains with the number of carbon atoms is more marked toward BL Lac than in the dense dark cloud TMC-1. However, the same pattern of higher abundances for chains with an even number of carbon atoms with respect to chains with an odd number of carbon atoms is observed. %\LEt{Single-sentence paragraph. ***}

These observations show the potential of deep observations in the Q band for detecting complex species in the diffuse interstellar medium. Similar observations in the K band could allow for the detection of longer chains if a matched sensitivity on the continuum approaching 0.01 \% can be reached.

\begin{acknowledgements}
This work is based on observations carried out with the Yebes 40 m telescope (project number 24A023). The 40 m radio telescope at Yebes Observatory is operated by the Spanish Geographic Institute (IGN; Ministerio de Transportes y Movilidad Sostenible).
B.T. acknowledges Spanish Ministry of Science support from grants PID2022-137980NB-100 and PID2023-147545NB-100.
This research has made use of spectroscopic and collisional data from the EMAA database (https://emaa.osug.fr and https://dx.doi.org/10.17178/EMAA). EMAA is supported by the Observatoire des Sciences de l’Univers de Grenoble (OSUG). 
This work was supported by the program "Physique et Chimie du Milieu Interstellaire (PCMI)" funded by Centre National de la Recherche Scientifique (CNRS) and Centre National d'Etudes Spatiales (CNES). 
The National Radio Astronomy Observatory is operated by Associated Universities, Inc. under a cooperative agreement with the National Science Foundation.
\end{acknowledgements}

\bibliographystyle{aa} 
\bibliography{bllac2} 

\onecolumn
\begin{appendix}
\section{BL Lac continuum flux}
\label{app:cont}

\Figcont

\Figpol

\Figspec

Figure \ref{fig:cont} presents the detected brightness temperature during the observations for each of the spectral backends. 
V-polarization measurements are shown in blue and H polarization in red. We detect a significant polarization fraction and variation of the flux across the spectral band and between the different observing days. The polarization fraction $p$ is illustrated in Fig \ref{fig:cont2}. It is defined as $p =\frac{T_V-T_H}{T_V+T_H}$, where $T_V$ and $T_H$ correspond to the
temperature in the V and H polarizations. We have also derived the brightness temperature spectral index for each continuum observation. The values are shown in Fig. \ref{fig:cont3}.

\FloatBarrier
\section{Additional tables}

\label{app:tables}

This appendix presents the tables of the detected lines, Table~\ref{tab:lines}, upper limits for selected molecular transitions, Table~\ref{tab:upper}, and the compilation abundances in diffuse and translucent clouds Table~\ref{tab:abun}.

\TabLine

\Tabupper

\Tabab

 \end{appendix}
\end{nolinenumbers}
\end{document}